\documentclass[useAMS,usenatbib]{aa}
\usepackage{graphicx}

\begin{document}

\title{The Elliptical Galaxy NGC\,5044: Stellar Population and Ionized Gas}

\author{M. G. Rickes\inst{1}, M. G. Pastoriza\inst{1} \and Ch. Bonatto\inst{1} }

\offprints{Ch. Bonatto - charles@if.ufrgs.br}

\institute{Universidade Federal do Rio Grande do Sul (UFRGS), Instituto de F\'\i sica,
 Departamento de Astronomia\\
\email{maurogr@if.ufrgs.br; mgp@if.ufrgs.br and charles@if.ufrgs.br} }

\date{Received --; accepted --}

\abstract{In this work we investigate the stellar population, metallicity distribution 
and ionized gas in the elliptical galaxy NGC\,5044, using long-slit spectroscopy and a 
stellar population synthesis method. We found differences in the slope of metal-line 
profiles along the galaxy which suggests an enhancement of $\alpha$ elements, particularly 
towards the central region.   The stellar population synthesis showed that the component with  
$[Z/Z_\odot]\sim0.0$ dominates the $\lambda5870$\,\AA\ flux in the central region of NGC\,5044, 
contributing with $\sim42\%$ of the total flux, while in the external  
regions the contribution decreases to $\sim8.0\%$.
The component with $[Z/Z_\odot]\sim-0.4$ contributes with $\sim32\%$ in the central 
region, and $\sim55\,\%$ in the external regions. The component with $[Z/Z_\odot]\sim-1.1$ 
contributes with $\sim26\,\%$ in the central region, and $\sim 37\,\%$
in the external regions.  The three components have $\sim\,10^{10}$ years.
The presence of a non-thermal ionization source, such as a low-luminosity AGN and/or shock 
ionization, is implied by the large values of the ratio $\frac{\ion{[N}{ii]}}{H\alpha}$ observed 
in all sampled regions.  However, the emission lines observed in the external regions indicate 
the presence of an additional ionization source, probably hot, post-AGB stars. 
}

\titlerunning{Stellar population and ionized gas in NGC\,5044}

\authorrunning{M.G. Rickes et al.} 

\maketitle

\section{Introduction}

A great effort has been made in the last few years to understand how early-type galaxies 
form and evolve. The relatively recent discoveries that ellipticals are not merely a one-parameter
family and that many ellipticals show signatures of interaction with the environment suggest that 
these galaxies may have different star formation histories, with stellar populations differing in 
metallicity and/or age (Worthey, Faber \& Gonzales 1992).

Crucial information on the above issues has been derived from metal line-strength indices
(e.g. Davies et al. 1987) and their radial variation inside the galaxies. The $Mg2$ line-strength
distribution in early-type galaxies, for example, can vary considerably, ranging from essentially
featureless to structured profiles showing, e.g., changes of slope possibly associated with kinematically
decoupled cores, or anomalies in the stellar population of some ellipticals  (Carollo et al. 1993).

NGC\,5044 is the central and brightest member of a rich group that contains many dE dwarf members 
and a few Im and Sm dwarf candidates. The mean radial velocity of the group is $\sim2048$ 
km\,s$^{-1}$ (Ferguson \& Sandage 1990). This galaxy presents a very bright ionized gas emission 
in the form of extended filaments up to $40\arcsec$ from the center, being larger in the southern 
part of the galaxy (Macchetto et al. 1996; Goudfrooij 1991). The dust distribution has an irregular 
morphology, concentrated in the inner $10\arcsec$, where two central dark clouds can be seen (Ferrari et 
al. 1999). NGC\,5044 has also been  detected by IRAS at 60 and $100\mu\,m$ and also at 6, 12, and 
$18\mu\,m$ by ISO (Ferrari et al. 2002). Its IRAS luminosity is $14\times 10^{8}\,L_\odot$. The galaxy 
has a nuclear radio source and extended X-ray emission (Fabbiano et al. 1992) and shows a very complex 
gas and stellar kinematics. The gas velocity profile is very irregular, with many  humps and dips; 
its radial velocity is systematically blueshifted with respect to the stellar systemic velocity by 
$\sim60 - 100$ km\,$\mathrm{s^{-1}}$.  Furthermore, the stellar velocity curves observed on a fairly 
large area near the galaxy center counter-rotate with respect to the outer regions. Very slow or no
stellar rotation is detected along the galaxy's minor axis. The velocity dispersion of the gas 
peaks at $\sim200 - 230$ km\,s$^{-1}$ in the center, decreasing rapidly outwards (Caon et al. 2000).
The peculiar kinematics and gas morphology described above may be signatures either of interactions
of NGC\,5044 with the environment or a post-merger event.

The goal of this paper is to investigate the stellar population, metallicity distribution and the 
presence of ionized gas in NGC\,5044, which are fundamental parameters to understand the formation 
and evolution of this galaxy.

This paper is organized as follows: Sect.~2 presents the observations and data reduction; 
Sect.~3 deals with the equivalent width analysis of the absorption lines and their radial 
dependence; Sect.~4 describes the stellar population synthesis, whose results are discussed 
in Sect.~5; Sect.~6  presents an analysis of the emission lines and the nature of the ionization 
source. Finally, in Sect.~7 a general discussion of the results and conclusions will be drawn.

\section{Observations and spectra extraction}

NGC\,5044 is a bright southern elliptical galaxy for which the general parameters are
given in Table~\ref{data}. 

\renewcommand{\tabcolsep}{7mm}
\begin{table}
\footnotesize
\begin{center}
\begin{tabular} {l c}
\hline 
\hline
Parameter                        &  NGC 5044                           \\
\hline
$\alpha$(J2000.0)                & $13^h15^m23.96^s$                   \\
$\delta$(J2000.0)                & $-16^\circ23\arcmin07.50\arcsec$    \\
Morphology                             & $E0$                          \\
$M_B$                            & $-21.92$                             \\
$B$                              & $11.83$                              \\
$E(B-V)$                         & $0.070$                              \\
$L_{H\alpha + [NII]}$ (erg\,$s^{-1}$)  & $1.28 \times 10^{41}$ \\ 
Radial velocity (km\,$s^{-1}$)   & $2704 \pm 33$        \\
Redshift                         & $0.00902 \pm 0.00011$ \\

Diameters (arcmin)               & $3.0\times3.0$        \\
$H_0$ (km\,$s^{-1}$ $Mpc^{-1}$)  & $75$                  \\                    

\hline
\end{tabular}
\end{center}
\caption{Data obtained at the NASA/IPAC Extragalactic Database (NED) which is operated by the Jet
Propulsion Laboratory, California Institute of Technology, under contract with the National Aeronautics
and Space Administration. }
\label{data}
\end{table}

\begin{figure}
\resizebox{\hsize}{!}{\includegraphics{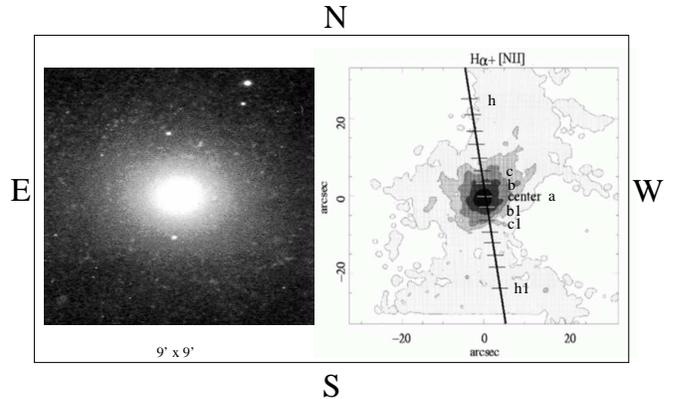}}
\caption{NGC\,5044. Left panel - V band image. Right panel - slit positioning and 
spatial extraction on an $\rm H\alpha$ image. Darker gray corresponds to larger flux.}  
\label{fig1}
\end{figure}

The star formation history, metallicity distribution and emission-gas properties
of NGC\,5044 are studied using long-slit spectroscopy. The observations were  obtained 
with the ESO 3.6m telescope at La Silla, Chile, equipped with EFOSC1. The spectra cover 
the range $\lambda\lambda5100 - 6800$\,\AA\ with a 3.6\AA/pixel resolution. The spatial 
scale of the observational configuration is $0.6\arcsec\rm\,pixel^{-1}$; the slit length 
was $3.1\arcmin$ while its width was fixed at $1.5\arcsec$ corresponding approximately to 
the average seeing; two 2-dimensional spectra were obtained with the same exposition time 
(approximately 45 minutes) and  position angle $PA = 10^\circ$. The slit positioning on the 
galaxy is shown in Fig.~\ref{fig1} (right panel), in which the dashes mark the spatial 
location of each extracted spectrum. A final one-dimensional spectrum was obtained combining 
(by weighted average) the two extractions for each position marked in Fig.~\ref{fig1}.

\begin{figure}
\resizebox{\hsize}{!}{\includegraphics{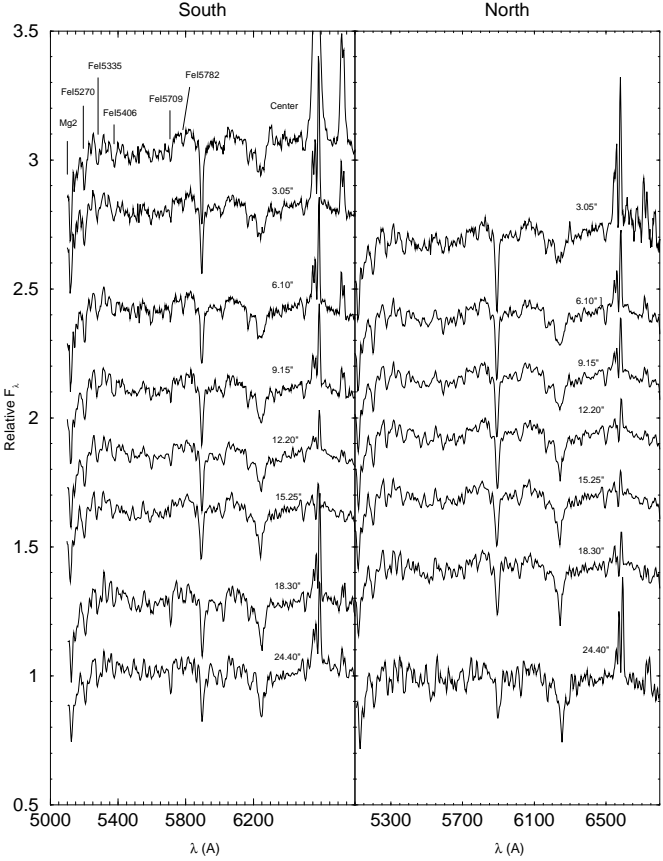}}
\caption{Spatial extractions for the South (left panel) and North (right panel) directions 
normalized at $\lambda5870$\,\AA. The distance to the galaxy center is indicated on each 
spectrum. Except for the bottom ones, the spectra have been shifted by an arbitrary constant, 
for clarity purposes}
\label{fig2}
\end{figure}

Details on the extracted spectra are given in Table~\ref{extracoes}. The spectra have been
extracted within different areas in order to increase the signal to noise ratio (S/N), since 
more external regions have lower flux. Each spectrum was corrected for radial velocity 
($V_r = 2704$\,km\,s$^{-1}$) and Galactic extinction ($A_V = 0.07$). The final average spectra,
normalized at $\lambda5870$\AA\ are shown in Fig.~\ref{fig2}, separately for the north 
and south directions on the galaxy.

%__________________________________________________ One column table

\renewcommand{\tabcolsep}{4.5mm}	
\begin{table}[h]
\footnotesize
\caption{Spatial extractions}
\begin{tabular} {r|c|c|c}
\hline
\hline
Spectra & R      & $R/R_{eff}$ & Extraction area \\
          & (kpc)  &             & ${\rm (kpc)}^2$ \\
\hline
centro    & $0.00$ & $0 $        & $0.06$ \\
bN,bS     & $0.53$ & $11$        & $0.06$ \\
cN,cS     & $1.06$ & $21$        & $0.13$ \\
dN,dS     & $1.59$ & $32$        & $0.19$ \\
eN,eS     & $2.12$ & $43$        & $0.25$ \\
fN,fS     & $2.66$ & $53$        & $0.30$ \\
gN,gS     & $3.19$ & $64$        & $0.30$ \\
hN,hS     & $4.26$ & $85$        & $0.58$ \\
\hline
\hline
\end{tabular}
\begin{list} {Table Notes.}
\item Column 2: Distance of the extraction region to the galaxy center. N and S
correspond to the north and south directions, respectively.
\end{list}
\label{extracoes}
\end{table}

%
%                                                One column figure
%----------------------------------------------------------- S_vib

\section{Equivalent width and radial dependence}

Elliptical galaxies present systematic variations on the line strength indices \ion{Mg}{2} and 
\ion{Fe}{i},
either from the center to the external regions of a given galaxy or limited to the central regions of
different galaxies (cf. Gonz\'ales 1993; Carollo \& Danziger 1994a,b; Carollo et al. 1993;
Davies et al. 1993; Fisher et al. 1995, 1996). Furthermore, the gradients in \ion{Mg}{2} and \ion{Fe}{i} 
have 
different slopes, suggesting an enhancement of Mg ($\alpha$ elements in general, which are associated 
with the Super-Novae type I rate) with respect to Fe toward the center of these galaxies. The inferred 
degree of enhancement seems to increase from dwarfs to massive ellipticals (see Faber et al. 1992, 
Worthey et al. 1994 and Matteucci 1997).

The above arguments suggest the occurrence of changes in some fundamental properties of the
constituent stellar population. In order to investigate if NGC\,5044, which presents a peculiar 
stellar kinematics and external ionized gas filaments, has stellar population and metallicity 
properties similar to those of other elliptical galaxies, we first investigate the spatial 
distribution of metal-line strengths, and then perform a stellar population analysis.

The present spectra of NGC\,5044 cover the spectral range $\lambda\lambda\,5100 -6800$\AA\, 
which constrains the number of important absorption lines used in stellar population studies. 
However, in this spectral region the galaxy presents several neutral iron lines, e.g. 
\ion{Fe}{i}$_{\lambda5270}$, \ion{Fe}{i}$_{\lambda5335}$, \ion{Fe}{i}$_{\lambda5406}$, 
\ion{Fe}{i}$_{\lambda5709}$ and \ion{Fe}{i}$_{\lambda5782}$, as well as the \ion{Mg}{2} band. 
These absorption features present strong dependence on the age and metallicity of the underlying 
stellar population (Bica \& Alloin 1986).

Previous to the measurements, all spectra have been normalized at $\lambda5870$\AA.
The continuum tracing and spectral windows of the absorption lines used in the present paper are
those defined by the Lick system (Faber et al. 1985), which are reproduced in Table~\ref{tablick}.
Each equivalent width (EW) and the \ion{Mg}{2} index have been measured three times taking into
account the uncertainties in the continuum level definition. This procedure allowed us to estimate 
the average value and corresponding standard deviation for each measurement.

Although \ion{Na}{i}$_{\lambda5895}$ is the strongest absorption feature present in our spectra,
it was not used in the analysis because it may be contaminated by dust concentrated in the central 
region of the galaxy (Ferrari et al. 1999). Besides the EWs and \ion{Mg}{2} index, we measured as 
well continuum points at 5300\AA, 5313\AA, 5406\AA, 5800\AA, 5870\AA\ and 6630\AA, as defined in 
Bica \& Alloin (1986). These continuum points, normalized at $\lambda\,5870\,\AA$, are listed in 
Table~\ref{continuos}.

In order to compare our results for NGC\,5044 with other elliptical galaxies which present
similar radial gradients in the $\ion{Mg}{2}$ band and in $\ion{Fe}{i}_{5270}$ and $\ion{Fe}{i}_{5335}$
lines (e.g. Davies et al., 1987; Carollo et al. 1993), we first corrected these features for the 
line broadening due to the internal stellar velocity dispersion.

The spectrum of the G giant star HR\,5333 was used to estimate the velocity dispersion correction.
HR\,5333 was assumed to have zero intrinsic velocity and its spectrum was broadened with a
series of gaussian filters with a velocity dispersion $\sigma$ varying from 
$\mathrm{0 - 300\,km\,s^{-1}}$, in steps of $\mathrm{10\,\,km\,s^{-1}}$.

For each absorption feature considered we calculate an empirical correction index ${\rm C(\sigma)}$, 
so that $C(\sigma)_{Mg2}=\frac{Mg2(0)}{Mg2(\sigma)}$, and 
$C(\sigma)_{FeI} =\frac{W_\lambda(0)}{W_\lambda(\sigma)}$.  
The velocity dispersion in NGC\,5044 varies from $\sim 260$ in the center to $\sim 230\,km\,s^{-1}$ in 
the external regions (Caon et al. 2000). We have calculated the correction index  $C(\sigma)$ for 
each marked position in Fig.~\ref{fig1} using Caon et al.'s (2000) $\sigma$ values. Our 
correction factors vary from 1.23 to 1.27 for the \ion{Mg}{2} index and from 1.20 to 1.23 for 
the \ion{Fe}{I} lines, in the velocity range 230--260\,km s$^{-1}$. Our correction factors for
the \ion{Mg}{2} index are somewhat larger than those given by Davies et al. (1993), while 
those for \ion{Fe}{i}$_{\lambda5270}$ are similar.

\begin{figure}
\resizebox{\hsize}{!}{\includegraphics{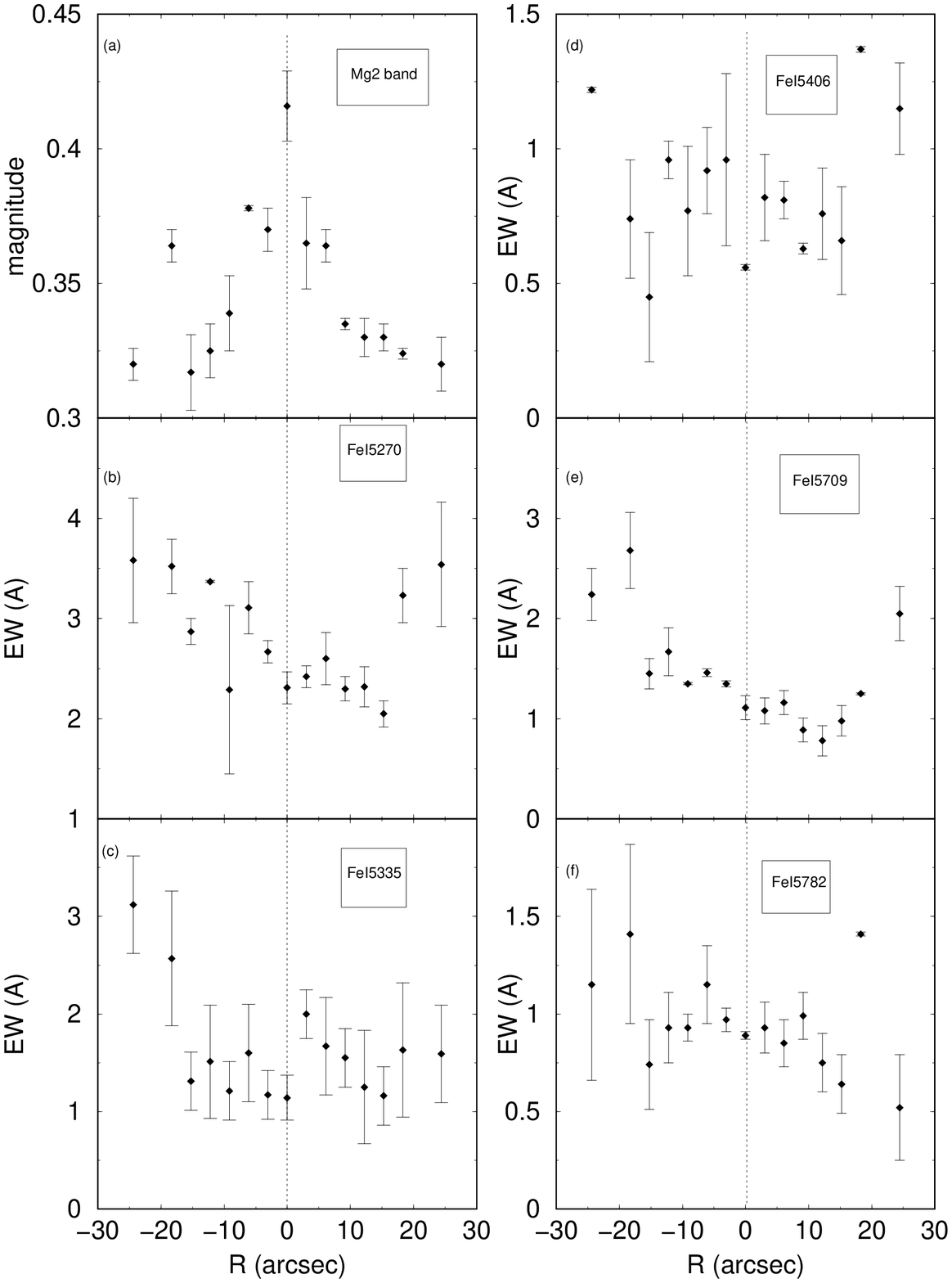}}
\caption{Spatial variation of absorption features in NGC\,5044.}
\label{fig3}
\end{figure}

As can be seen in Fig.~\ref{fig3}, except for \ion{Fe}{i}$_{\lambda5406}$ (panel(d)),  
the other absorption features present strong variations with distance to the center. In 
particular, the \ion{Mg}{2} band (panel (a)) clearly presents a positive gradient, i.e. 
increasing towards the center of the galaxy, whereas the other $\ion{Fe}{i}$ lines present 
evidence of negative gradients.

% ***************************** INDICES DE LICK ********************************

\renewcommand{\arraystretch}{0.8}
\renewcommand{\tabcolsep}{0.5mm}
\begin{table}
\footnotesize
\caption{Lick indices definitions}
\begin{tabular}{l c c c}
\hline
\hline
Index & Line/Band & $\Delta\,\lambda$ (\AA) &
AC (\AA) \\
\hline
$Mg2_{Band}$ & $Mg2$ & $5155.375$|$5197.875$ & $4896.375$|$4958.875$ \\ 
 & & & $5302.375$|$5367.375$ \\
\hline
$Fe_{\lambda5270}$  & \ion{Fe}{i} & $5247.375$|$5287.375$ & $5234.875$|$5249.875$ \\
 & & & $5287.375$|$5319.875$ \\
\hline

$Fe_{\lambda5335}$  & \ion{Fe}{i} & $5314.125$|$5354.125$ & $5306.625$|$5317.875$ \\
 & & & $5355.375$|$5365.375$ \\
\hline
$Fe_{\lambda5406}$  & \ion{Fe}{i} & $5390.250$|$5417.750$ & $5379.000$|$5390.250$ \\
 & & & $5417.750$|$5427.750$ \\
\hline
$Fe_{\lambda5709}$  & \ion{Fe}{i} & $5698.375$|$5722.125$ & $5674.625$|$5698.375$ \\
 & & & $5724.625$|$5738.375$ \\
\hline
$Fe_{\lambda5782}$  & \ion{Fe}{i} & $5778.375$|$5798.375$ & $5767.125$|$5777.125$ \\
 & & & $5799.625$|$5813.375$ \\
\hline
\hline
\end{tabular}
\begin{list} {Table Notes.}
\item Column 3: Spectral window; Column 4: Adjacent Continuum.
\end{list}
\label{tablick}
\end{table}

%****************************** LARGURAS EQUIVALENTES *****************************************
\renewcommand{\tabcolsep}{3.1mm}
\begin{table*}
\footnotesize
\caption{\ion{Mg}{2} index and EWs of the \ion{Mg}{2} and \ion{Fe}{i} lines.}
\begin{tabular} {l|cclccccc}
\hline
\hline
&mag&&\multicolumn{5}{c}{EW (\AA)} \\
\cline{2-2}\cline{4-9}\\
R ($\arcsec$)& ${\rm \ion{Mg}{2}}_{index}$ & & ${\rm \ion{Mg}{2}}_{band}$ & \ion{Fe}{i}$_{\lambda5270}$ &
\ion{Fe}{i}$_{\lambda5335}$
& \ion{Fe}{i}$_{\lambda5406}$   & \ion{Fe}{i}$_{\lambda5709}$ &
\ion{Fe}{i}$_{\lambda5782}$ \\
\hline
$ 0$     &  $ 0.416  \pm 0.013$ && $13.92 \pm 0.51 $ & $2.31 \pm 0.16$ & $1.14 \pm 0.23$ & $0.56 
\pm 0.01$ 
& $1.11 \pm 0.12$ & $0.89 \pm 0.02$ \\
$3.05S $ &  $ 0.370 \pm  0.008$ && $13.33 \pm 0.38 $ & $2.67 \pm 0.11$ & $1.17 \pm 0.25$ & $0.96 

\pm 0.32$ 
& $1.35 \pm 0.03$ & $0.97 \pm 0.06$ \\
$6.10S $ &  $ 0.378 \pm  0.001$ && $11.85 \pm 0.62 $ & $3.11 \pm 0.26$ & $1.60 \pm 0.50$ & $0.92 
\pm 0.16$ 
& $1.46 \pm 0.04$ & $1.15 \pm 0.20$ \\
$9.15S $ &  $ 0.339 \pm  0.014$ && $11.72 \pm 0.70 $ & $2.29 \pm 0.64$ & $1.21 \pm 0.30$ & $0.77 
\pm 0.24$ 
& $1.35 \pm 0.01$ & $0.93 \pm 0.07$ \\
$12.20S$ &  $ 0.325 \pm  0.010$ && $11.20 \pm 0.42$  & $3.37 \pm 0.01$ & $1.51 \pm 0.58$ & $0.96 
\pm 0.07$ 
& $1.67 \pm 0.24$ & $0.93 \pm 0.18$ \\
$15.25S$ &  $ 0.317 \pm  0.014$ && $11.69 \pm 0.70$  & $2.87 \pm 0.13$ & $1.31 \pm 0.30$ & $0.45 
\pm 0.24$ 
& $1.45 \pm 0.15$ & $0.74 \pm 0.23$ \\
$18.30S$ &  $ 0.364 \pm  0.006$ && $11.65 \pm 0.51$  & $3.52 \pm 0.27$ & $2.57 \pm 0.69$ & $0.74 
\pm 0.22$ 
& $2.68 \pm 0.38$ & $1.41 \pm 0.46$ \\
$24.40S$ &  $ 0.320 \pm  0.006$ && $11.23 \pm 0.36$  & $3.58 \pm 0.62$ & $3.12 \pm 0.50$ & $1.22 
\pm 0.01$ 
& $2.24 \pm 0.26$ & $1.15 \pm 0.49$ \\
\hline

$3.05N $ & $ 0.365  \pm  0.017$ && $13.29 \pm 0.38$  & $1.99 \pm 0.11$ & $2.00 \pm 0.25$ & $0.82 
\pm 0.16$ 
& $1.08 \pm 0.13$ & $0.93 \pm 0.13$ \\
$6.10N $ & $ 0.364  \pm  0.006$ && $12.62 \pm 0.62$  & $2.13 \pm 0.26$ & $1.67 \pm 0.50$ & $0.81 
\pm 0.07$ 
& $1.16 \pm 0.12$ & $0.85 \pm 0.12$ \\
$9.15N $ & $ 0.335  \pm  0.002$ && $11.47 \pm 0.70$  & $1.92 \pm 0.12$ & $1.55 \pm 0.30$ & $0.63 
\pm 0.02$ 
& $0.89 \pm 0.12$ & $0.99 \pm 0.12$ \\
$12.20N$ & $ 0.330  \pm  0.007$ && $11.71 \pm 0.22$  & $1.93 \pm 0.20$ & $1.25 \pm 0.58$ & $0.76 
\pm 0.17$ 
& $0.78 \pm 0.15$ & $0.75 \pm 0.15$ \\
$15.25N$ & $ 0.330  \pm  0.005$ && $11.54 \pm 0.17$  & $2.21 \pm 0.13$ & $1.16 \pm 0.30$ & $0.66 
\pm 0.20$ 
& $0.98 \pm 0.15$ & $0.64 \pm 0.15$ \\
$18.30N$ & $ 0.324  \pm  0.002$ && $12.33 \pm 0.51$  & $2.69 \pm 0.27$ & $1.63 \pm 0.69$ & $1.37 
\pm 0.01$ 
& $1.25 \pm 0.01$ & $1.41 \pm 0.01$ \\
$24.40N$ & $ 0.320  \pm  0.010$ && $11.29 \pm 0.36$  & $2.92 \pm 0.62$ & $1.59 \pm 0.50$ & $1.15 
\pm 0.17$ 
& $2.05 \pm 0.27$ & $0.52 \pm 0.27$ \\
\hline
\end{tabular}
\begin{list} {Table Notes.}
\item Column 1: Distance of the extraction to the galactic center. N and S
correspond to the north and south directions, respectively.
\end{list}
\label{largura}
\end{table*}

%************************ PONTOS DE CONTINUO **********************************

\renewcommand{\tabcolsep}{5.4mm} 
\begin{table*}
\caption{Continuum points}
\begin{tabular} {l|c c c c c c }
\hline
\multicolumn{7}{c}{${\rm C}_{\lambda}$/${\rm C}_{5870}$} \\
\hline
\hline
& & & & & & \\
R ($\arcsec$)& $\lambda_{5300}$ & $\lambda_{5313}$ & $\lambda_{5546}$ &
$\lambda_{5800}$ & $\lambda_{5822}$ & $\lambda_{6630}$ \\
\hline
 $0$ & $0.95 \pm 0.03$ & $1.02 \pm 0.07$ & $0.99 \pm 0.02$ & $1.02 \pm
0.05$ & $1.02 \pm 0.05$ & $0.98 \pm 0.12$ \\ 
 $3.05S$ & $0.99 \pm 0.04$ & $1.00 \pm 0.01$ & $0.96 \pm 0.05$ & $1.00 \pm
0.01$ & $1.00 \pm 0.02$ & $0.95 \pm 0.05$ \\ 
 $6.10S$ & $1.03 \pm 0.05$ & $1.03 \pm 0.02$ & $0.98 \pm 0.05$ & $1.01 \pm
0.02$ & $1.02 \pm 0.04$ & $0.95 \pm 0.06$ \\ 
$9.15S$ & $1.01 \pm 0.05$ & $1.02 \pm 0.01$ & $0.96 \pm 0.06$ & $0.99 \pm
0.02$ & $1.00 \pm 0.02$ & $0.93 \pm 0.05$ \\
 $12.20S$ & $1.01 \pm 0.05$ & $1.01 \pm 0.01$ & $0.97 \pm 0.05$ & $0.99 \pm
0.02$ & $0.99 \pm 0.01$ & $0.94 \pm 0.06$ \\ 
 $15.25S$ & $1.04 \pm 0.05$ & $1.04 \pm 0.01$ & $0.98 \pm 0.05$ & $1.00 \pm
0.01$ & $1.00 \pm 0.02$ & $0.92 \pm 0.06$ \\ 
 $18.30S$ & $1.01 \pm 0.04$ & $1.02 \pm 0.02$ & $0.98 \pm 0.04$ & $1.02 \pm
0.03$ & $1.01 \pm 0.03$ & $0.96 \pm 0.07$ \\
 $24.40S$ & $1.02 \pm 0.05$ & $1.03 \pm 0.02$ & $1.00 \pm 0.02$ & $1.01 \pm
0.02$ & $1.01 \pm 0.02$ & $0.94 \pm 0.07$ \\
\hline
 $3.05N$ & $0.99 \pm 0.06$ & $0.99 \pm 0.01$ & $0.95 \pm 0.05$ & $0.99 \pm
0.01$ & $1.01 \pm 0.03$ & $0.95 \pm 0.05$ \\ 
 $6.10N$  & $0.99 \pm 0.04$ & $1.00 \pm 0.02$ & $0.95 \pm 0.06$ & $1.00 \pm
0.01$ & $1.01 \pm 0.03$ & $0.94 \pm 0.05$ \\ 
   $9.15N$  & $1.00 \pm 0.05$ & $1.00 \pm 0.02$ & $0.97 \pm 0.04$ & $1.00 \pm
0.01$ & $0.99 \pm 0.02$ & $0.94 \pm 0.05$ \\
 $12.20N$  & $1.00 \pm 0.04$ & $1.00 \pm 0.03$ & $0.97 \pm 0.01$ & $0.99 \pm
0.03$ & $1.01 \pm 0.02$ & $0.95 \pm 0.06$ \\ 
 $15.25N$  & $1.02 \pm 0.05$ & $1.02 \pm 0.01$ & $0.98 \pm 0.05$ & $1.01
\pm 0.01$ & $1.01 \pm 0.01$ & $0.95 \pm 0.06$ \\ 
 $18.30N$  & $1.00 \pm 0.06$ & $1.00 \pm 0.02$ & $0.97 \pm 0.02$ & $1.00 \pm
0.01$ & $1.00 \pm 0.02$ & $0.94 \pm 0.05$ \\
 $24.40N$  & $1.01 \pm 0.05$ & $1.01 \pm 0.01$ & $1.01 \pm 0.04$ & $1.00 \pm
0.01$ & $0.99 \pm 0.01$ & $0.93 \pm 0.04$ \\
\hline
\end{tabular}
\begin{list} {Table Notes.}
\item  N and S correspond to north and south respectively.
\end{list}
\label{continuos}
\end{table*}

%*************** GRAFICO COM OS FATORES DE CORRECAO POR DISP. DE VELOCIDADE ****

%\begin{figure}
%   \centering
%   \resizebox{\hsize}{!}{\includegraphics{features_sigma.eps}}
%     \caption{Dependence of the correction factor on the velocity dispersion .
%              }
%   \label{fatorcorrecao}
%   \end{figure}
%*******************************************************************************

\subsection{Metallicity gradient}

\begin{figure}
\resizebox{\hsize}{!}{\includegraphics{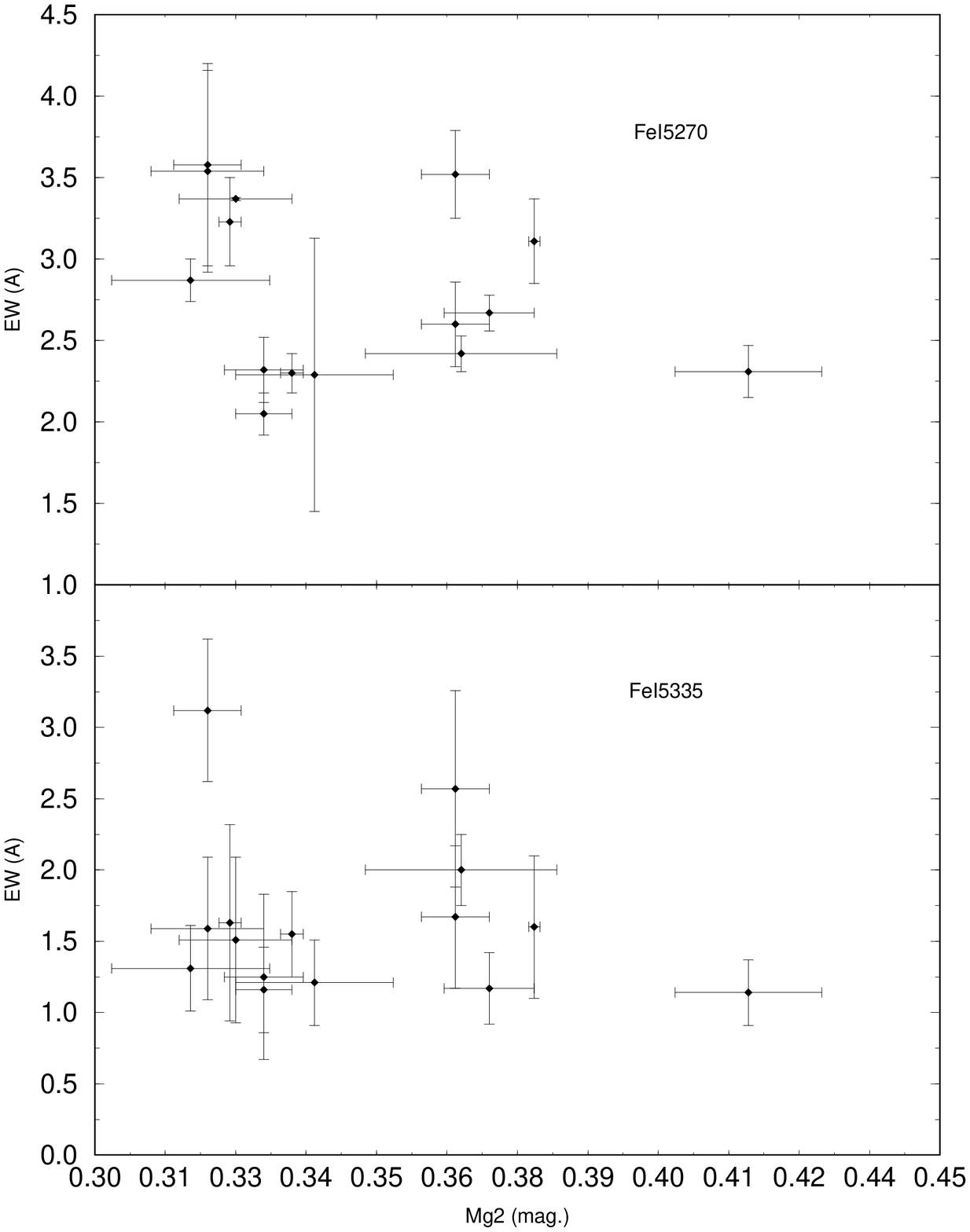}}
\caption{\ion{Mg}{2} index vs. \ion{Fe}{i}$_{5270}$ and \ion{Fe}{i}$_{5335}$ line strength.}
\label{fig4}
\end{figure}

The radial distributions of the \ion{Mg}{2} index and those of the EWs of \ion{Fe}{i}$_{5270}$, 
\ion{Fe}{i}$_{5335}$, \ion{Fe}{i}$_{5406}$, \ion{Fe}{i}$_{5709}$ and \ion{Fe}{i}$_{5782}$, 
corrected for velocity dispersion, are shown in Fig.~\ref{gradlarg02}. The corrected \ion{Mg}{2} 
index (panel(a)) presents an enhancement of the gradient which decreases from $0.42\,\,mag$ in 
the center to $\sim\,0.32\,\,mag$ in the external regions of the galaxy. On the other hand, the 
corrected EW of $\ion{Fe}{i}_{5270}$ (panel(b)) increases from 2.30\,\AA\ in the center to 
$\sim$3.57\,\AA\,\, in the external regions. This result is in reasonable agreement with Carollo 
et al. (1993) who found no correlation of the \ion{Mg}{2} index with either \ion{Fe}{i}$_{5270}$ 
and \ion{Fe}{i}$_{5335}$. In Fig.~\ref{fig4} we plot the velocity dispersion corrected values 
of the \ion{Mg}{2} index {\it vs.} the EWs of \ion{Fe}{i}$_{5270}$ and \ion{Fe}{i}$_{5335}$, in 
which we see no correlation between these features. The different slope of the \ion{Mg}{2} gradient 
relative to those of \ion{Fe}{i}$_{5270}$ and \ion{Fe}{i}$_{5335}$ can be accounted for by an 
enhancement of $\alpha$-elements in general, Mg in particular (Faber et al. 1992 and Worthey et 
al. 1994). In NGC\,5044 this effect is more intense in the center than in the outer regions. 

The dependence of metallicity gradients on fundamental parameters was investigated by 
Carollo et al. (1993) who found that the \ion{Mg}{2} gradient shows a bimodal trend. For objects 
with masses lower than $10^{11}\,M_\odot$, the \ion{Mg}{2} gradient slope correlates with galaxy 
mass, whereas for objects more massive than $10^{11}\,M_\odot$, no such correlation appears. 

In order to compare our results with the data of Carollo et al. (1993), first we estimate the 
mass of NGC\,5044 using Poveda's approximation, $M_{tot} = 0.9\,\frac{\sigma^2\,R_e}{G} =
3.34\times10^8\,M_\odot$, where $G$ is the gravitational constant, $R_e = 28.53\arcsec$ is the 
effective radius and $\sigma = 256\,km\,s^{-1}$ is the central stellar velocity dispersion. 
According to Fig.~11 of Carollo et al. (1993), the calculated mass value of NGC\,5044 together 
with the slope of the \ion{Mg}{2} gradient, suggest that this elliptical galaxy was formed by a 
monolitic collapse. 

On the other hand, Kobayashi et al. (1999) have examined line-strength gradients in 80 elliptical 
galaxies.  They found typical gradients $\delta[Fe/H]/\delta log(r)\sim-0.3$, which are flatter than 
the ones predicted by monolithic collapse simulations. In this paper we 
found, using the same method, $\delta[Fe/H]/\delta log(r) = -0.31$ which is very close to the 
value obtained by Kobayashi et al. (1999).

\subsection{[Fe/H] abundance}

In order to estimate the [Fe/H] abundance in NGC\,5044, we use the calibrations \ion{Mg}{2}
{\it vs.} [Fe/H], \ion{Fe}{i}$_{5270}$ {\it vs.} [Fe/H] and \ion{Fe}{i}$_{5335}$ {\it vs.} [Fe/H] 
presented by Worthey et al. (1992) which are based on models of galactic globular clusters with 
18 Gyrs of age. From Figure~1 of Worthey et al. (1992) we found that for our corrected  \ion{Mg}{2} 
index, [Fe/H]= 0.5 and 0.17 in the central and external regions, respectively. In the same way, 
using \ion{Fe}{i}$_{5270}$ we found [Fe/H] = $-$0.5, 0.2; and for \ion{Fe}{i}$_{5335}$, we found 
[Fe/H] = $-$1.0, $-$0.5 in the central and external regions, respectively. Therefore, we found 
an [Fe/H] abundance above solar using as metallicity indicator the \ion{Mg}{2} index, while 
[Fe/H] is lower than solar for \ion{Fe}{i}$_{5270}$ and \ion{Fe}{i}$_{5335}$ lines; in addition 
Mg is above solar in NGC\,5044.

%*********************** GRADIENTES DE Mg2, FeI5270 E FeI5335 ********************

%\begin{figure}
%   \centering
%   \resizebox{\hsize}{!}{\includegraphics{gradfeatures_correct.eps}}
%     \caption{Mg2 index and EWs of $\ion{Fe}{i}_{5270}$ and  $\ion{Fe}{i}_{5335}$ 
%corrected for velocity dispersion.
%              }
%   \label{gradfeaturesc}
%   \end{figure}

%**********************************************************************************

%********************************** GRAFICO DE Mg2 VS. FeI ***********************
%**********************************************************************************

\section{Stellar population synthesis}

For a good understanding of some properties of NGC\,5044, such as the presence of central gas
emission, a precise determination of the star formation process is necessary. Therefore, the 
age of the constituent stars in NGC\,5044 is an important parameter to be determined.
        
The integrated spectrum of a given galaxy contains significant information on its stellar content
and chemical enrichment (Bica \& Alloin, 1985). In principle, this information together with a 
stellar population synthesis method, can be used to determine the star formation history (Bica 
\& Alloin, 1986). In this work we employ the stellar population synthesis method developed by 
Bica (1988) which is based on integrated spectra of star clusters and \ion{H}{ii} regions, 
characterized by different ages and metallicities. In the present case, due to the small number 
of observational constraints (EWs and continuum points), we use three old components with  
different metallicities in order to minimize the degeneracy between age and metallicity of 
the different  population components. These metallicities components (templates) are: $G1\, 
\rightarrow\,[Z/Z_{\odot}]\sim0.0$, $G2\,\rightarrow\,[Z/Z_{\odot}]\sim-0.4$ and $G3\, 
\rightarrow\,[Z/Z_{\odot}]\sim-1.1$, the three with age $10^{10}$ years. 

\begin{figure}
%   \centering
\resizebox{\hsize}{!}{\includegraphics{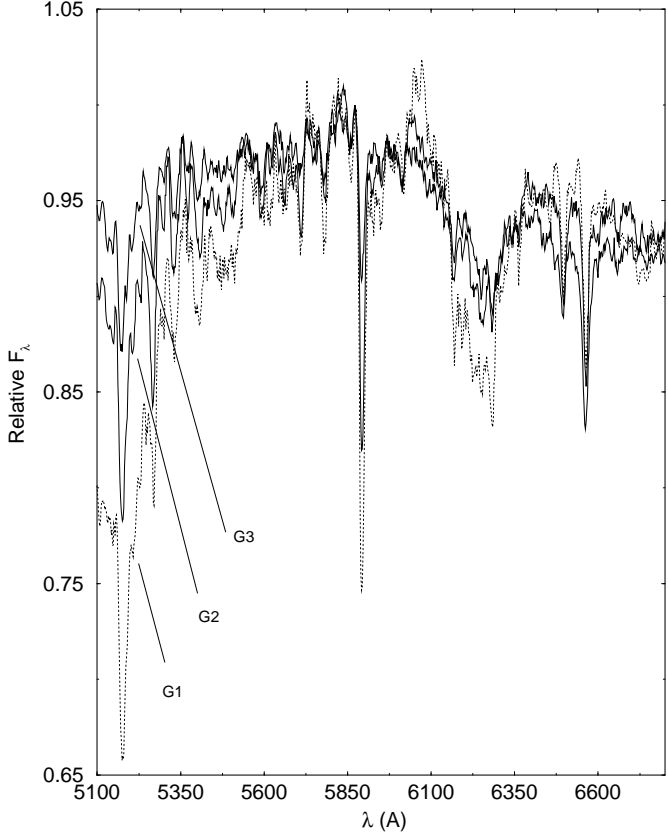}}
\caption{Population templates (age components) used in the synthesis, normalized 
at $\lambda5870$\,\AA.}
\label{fig5}
\end{figure}

Alternatively, we have also performed a stellar population synthesis with a set of three 
components with similar (solar) metallicities and different ages: G1 ($10^{10}$\,yr),
Y3 ($10^{8}$\,yr) and RHII ($10^{6}$\,yr). We found that the spectra presented in 
Fig.~\ref{fig2} are better fitted by a range in metallicities than a range in ages.
The base elements that we use are taken from the star cluster population templates described 
in Bica \& Alloin (1986), and are presented in Fig.~\ref{fig5}. The corresponding EWs 
and continuum points for the base elements have been measured similarly as those for NGC\,5044 
and are given in Tables~\ref{largbase} and \ref{contbase}, respectively. We remind that, for 
the stellar population analysis we use the EW of \ion{Mg}{2}, which is given in 
column (3) of Table~\ref{largura}.

Basically the algorithm uses the selected EWs and pivot points for the continuum
measured in a given spectrum and compares them to those of a model computed from
a base of simple stellar population elements. The algorithm is not a minimization
procedure, instead it generates all combinations of the base elements according to
a given flux contribution step. The code also successively dereddens galaxy input
continuum points and compares them against a given base model. Every combination
is subsequently tested against a set of windows of maximum allowed difference
between the observed and the resulting features obtained. Finally, the solutions
satisfying all feature windows are averaged out, and this average is adopted as
the final synthesis solution.

Initially we used a 10\% step for testing flux contribution at $\lambda5870$\,\AA\ generating 
about $\sim5\,200$ combinations for each assumed $E(B-V)_{\rm{i}}$.
Reddenings were tested in the range 0.00\,$\leq$\,E(B-V)$_{\rm{i}}$\,$\leq$\,0.50 with
a step of 0.01. Thus, in total, $\sim260\,000$ combinations are tested for each
extraction, and the number of possible solutions amounts to less than 1\%. After
probing as above the space of combinations, we calculate the solution with finer
steps of 5\%, 2\% and finally 1\%. We tested the Galactic (Seaton 1979), LMC
(Fitzpatrick 1986) and SMC (Pr\'evot et al. 1984) reddening laws for the
synthesis of each extraction, and concluded that the Galactic law applies in all cases.

%
%*********************************************************************************
%
%
%*********************************************************************************
%
\begin{table}
\caption{Equivalent widths for the base}
\renewcommand{\tabcolsep}{0.35mm}
\small
%\tiny
\begin{tabular} {l|c|c c c c c c}
\hline
\multicolumn{7}{c}{EW (\AA)} \\
\hline
 & Metallicity & & & & & & \\

 & $[Z/Z_{\odot}]$ & $\ion{Mg}{2}_{band}$ & $\ion{Fe}{i}_{5270}$ &
$\ion{Fe}{i}_{5335}$ & $\ion{Fe}{i}_{5406}$ & $\ion{Fe}{i}_{5709}$ &
$\ion{Fe}{i}_{5782}$  \\
\hline
$G1$ & $\sim \,\, 0.0$ & $5.82$ & $2.77$ & $1.65$ & $1.50$ & $1.36$ & $1.14$ \\
$Y3$ & $\sim -0.4$ & $4.29$ & $2.99$ & $1.70$ & $1.46$ & $1.30$ & $0.80$ \\
$Y3$ & $\sim -1.1$ & $2.75$ & $1.59$ & $0.95$ & $0.77$ & $0.57$ & $0.52$  \\
\hline
\end{tabular}
%\begin{list} {Tabl}
%\item
%\end{list}
\label{largbase}
\end{table}
%
%*********************************************************************************
%

\renewcommand{\tabcolsep}{2.3mm}
\begin{table}
\caption{Continuum points for the base}

\begin{tabular} {c|c c c c c c c c}
\hline
\multicolumn{7}{c}{{${\rm C}_{\lambda}/C_{5870}$}} \\
\hline

 & $\lambda5300$ & $\lambda5313$ & $\lambda5546$ & $\lambda5800$ & $\lambda5822$ &
$\lambda6630$ \\
\hline
$G1$ & $0.91$ & $0.97$ & $1.01$ & $1.01$ & $1.00$ & $0.94$ \\
$G2$ & $0.97$ & $0.98$ & $0.99$ & $1.01$ & $1.00$ & $0.94$ \\
$G3$ & $0.97$ & $0.99$ & $0.99$ & $0.99$ & $1.00$ & $0.93$ \\
\hline
\end{tabular}
\label{contbase}
\end{table}

%**********************************************************************************
%
\section{Synthesis results}

The results of our synthesis are given in Table~\ref{tabcontrib}, where
we present the percentage contribution of each base element to the flux at
$\lambda5870$\,\AA. The values of E(B$-$V)$_{\rm{i}}$ are also given in the table.

The G1 component ($[Z/Z_\odot]\sim 0.0$) dominates the $\lambda5870$\,\AA\ flux in 
the central extractions of NGC\,5044. In the nucleus, the G1 component contributes with 
$\sim42\%$ of the total flux, while in the external regions the contribution decreases to 
$\sim8.0\%$.

The G2 component ($[Z/Z_\odot]\sim-0.4$) contributes with $\sim32\%$ in the central 
region, and $\sim55\,\%$ in the external regions ($24.40\arcsec$). The G3 component
($[Z/Z_\odot]\sim-1.1$) contributes with $\sim26\,\%$ in the central region, and $\sim 37\,\%$
in the external regions, as can be seen in Table~\ref{tabcontrib}. Flux fractions and the 
internal reddening as a function of distance to the center are shown in Fig.~\ref{fig7}.   
Notice that the spatial distribution of the internal reddening E(B$-$V)$_{\rm{i}}$ along the 
north and south directions (Table~\ref{tabcontrib} and panel (d) in Fig.~\ref{fig7}) does
not characterize a gradient.

\begin{figure}
\resizebox{\hsize}{!}{\includegraphics{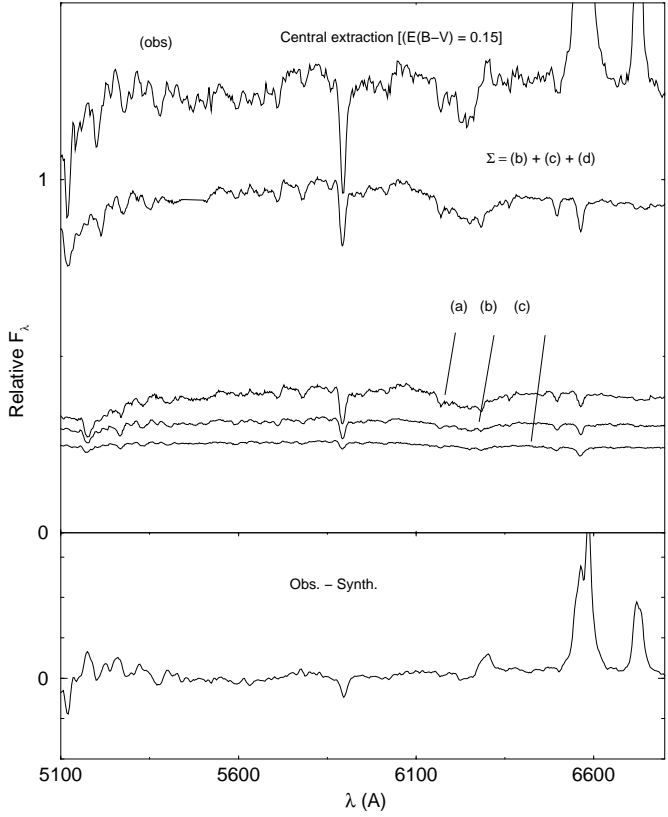}}
\caption{Stellar population synthesis of the central extraction. Top panel: (obs) - observed
spectrum corrected for reddening; (a) - G1 population template; (b) - G2 population template; 
(c) - G3 population template and ($\sum$) - synthesized spectrum.  Bottom panel:
pure emission spectrum. Spectrum (obs) has been shifted by a constant.}
\label{fig6}
\end{figure}

\begin{figure}
%   \centering
\resizebox{\hsize}{!}{\includegraphics{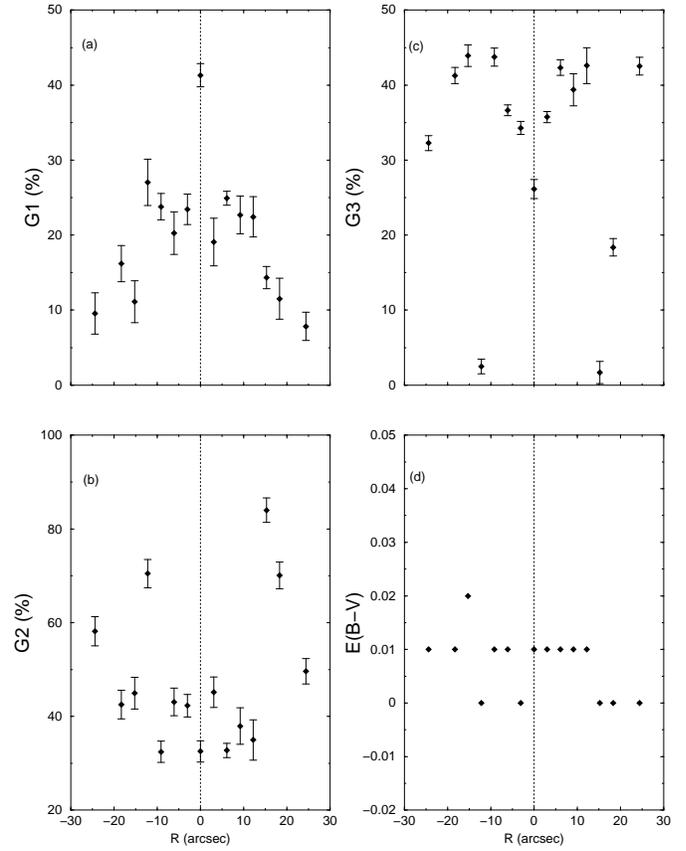}}
\caption{Synthesis results in flux fractions as a function of distance to the center -
Panels (a), (b) and (c); panel (d) - spatial distribution of the internal reddening.}
\label{fig7}
\end{figure}

The spectra representing the stellar population of NGC\,5044 have been constructed using the 
star cluster templates (Sect.~4) combined according to the proportions given by the synthesis. 
We illustrate this procedure in Fig.~\ref{fig6} in which we show the synthesis 
for the central extraction (top panel). The resulting pure emission spectrum, which is obtained 
after the subtraction of the population template, is shown in the bottom panel.

The goodness of the above stellar population synthesis method can be assessed by the residuals
both in the EWs and continuum points. As can be seen in Table~\ref{residualw}, most of the 
absorption features are reasonably well reproduced by the synthesis method. The residuals are 
almost equally distributed between positive and negative ones, probably indicating that NGC\,5044 has the 
same metallicity than that of the templates.  The continuum fit is quite good as well, as can be seen by the 
residuals in Table~\ref{residualc}.

%*********************************************************************************
%

\renewcommand{\arraystretch}{0.8}
\renewcommand{\tabcolsep}{1.0mm}
\begin{table}
\footnotesize
\caption{Synthesis results in terms of flux fractions}
\begin{tabular} {l|c c c c}
\multicolumn{5}{c}{\it{Flux fraction at $\lambda5870$ \AA }} \\
\hline
R ($\arcsec$) & G1 & G2 & G3 & $E(B-V)_i$ \\
 & (\%) & (\%) & (\%) & \\
\hline
$0.00    $ & $41.31 \pm 1.56$ & $32.53 \pm 2.30$ & $26.16 \pm 1.27$ & $0.01$ \\
$3.05S   $ & $23.43 \pm 2.04$ & $42.29 \pm 2.40$ & $34.29 \pm 0.88$ & $0.00$ \\
$6.10\,S $ & $20.27 \pm 2.85$ & $43.09 \pm 2.96$ & $36.64 \pm 0.73$ & $0.01$ \\
$9.15\,S $ & $23.79 \pm 1.76$ & $32.46 \pm 2.27$ & $43.75 \pm 1.2$ &  $0.01$ \\
$12.20\,S$ & $27.04 \pm 3.09$ & $70.48 \pm 3.01$ & $ 2.48 \pm 0.98$ & $0.00$ \\
$15.25\,S$ & $11.13 \pm 2.79$ & $44.95 \pm 3.40$ & $43.92 \pm 1.45$ & $0.02$ \\
$18.30\,S$ & $16.19 \pm 2.42$ & $42.53 \pm 3.03$ & $41.28 \pm 1.10$ & $0.01$ \\
$24.40\,S$ & $9.57  \pm 2.75$ & $58.16 \pm 3.14$ & $32.27 \pm 1.01$ & $0.01$ \\
\hline
$3.05\,N $ & $19.09 \pm 3.17$ & $45.17 \pm 3.25$ & $35.74 \pm 0.74$ & $0.01$ \\
$6.10\,N $ & $24.92 \pm 0.95$ & $32.75 \pm 1.53$ & $42.33 \pm 1.03$ & $0.01$ \\
$9.15\,N $ & $22.69 \pm 2.53$ & $37.93 \pm 3.89$ & $39.38 \pm 2.13$ & $0.01$ \\
$12.20\,N$ & $22.43 \pm 2.69$ & $34.97 \pm 4.29$ & $42.60 \pm 2.40$ & $0.01$ \\
$15.25\,N$ & $14.33 \pm 1.49$ & $84.00 \pm 2.58$ & $ 1.67 \pm 1.49$ & $0.00$ \\
$18.30\,N$ & $11.52 \pm 2.72$ & $70.10 \pm 2.84$ & $18.38 \pm 1.15$ & $0.00$ \\
$24.40\,N$ & $7.83  \pm 1.88$ & $49.64 \pm 2.69$ & $42.53 \pm 1.20$ & $0.00$ \\
\hline
\end{tabular}
\begin{list} {Table Notes.}
\item Column 1: S and N indicate south and north directions, respectively; Column~5:
internal reddening, corresponding to the Galactic law.
\end{list}
\label{tabcontrib}
\end{table}

%
%**********************************************************************************
%
\renewcommand{\tabcolsep}{1.0mm}
\begin{table}
%\footnotesize
\caption{Equivalent width residuals (\AA)}
\footnotesize
%\tiny
\begin{tabular} {l|r r r r r r}
%\multicolumn{7}{c}{}\\
\hline
\hline
R (\arcsec) & $\ion{Mg}{2}_{band}$ & $\ion{Fe}{i}_{5270}$ &
$\ion{Fe}{i}_{5335}$ & $\ion{Fe}{i}_{5406}$ & $\ion{Fe}{i}_{5709} $ &
$\ion{Fe}{i}_{5782}$ \\
\hline
$0.00$     &  $0.48 $ & $-0.21$ & $-0.34$ & $-0.73$ & $-0.02$ & $0.02$ \\
$3.05\,S$  &  $0.24 $ & $ 0.22$ & $-0.26$ & $-0.27$ & $ 0.29$ & $0.19$ \\
$6.10\,S$  &  $0.85 $ & $ 0.69$ & $ 0.19$ & $-0.29$ & $ 0.41$ & $0.38$ \\
$9.15\,S$  &  $0.80 $ & $-0.02$ & $-0.15$ & $-0.40$ & $ 0.35$ & $0.17$ \\
$12.20\,S$ &  $0.57 $ & $ 0.47$ & $-0.16$ & $-0.49$ & $ 0.37$ & $0.04$ \\
$15.25\,S$ &  $0.94 $ & $ 0.53$ & $-0.05$ & $-0.71$ & $ 0.46$ & $0.02$ \\
$18.30\,S$ &  $0.80 $ & $ 1.16$ & $ 1.19$ & $-0.44$ & $ 1.67$ & $0.66$ \\
$24.40\,S$ &  $0.32 $ & $ 1.07$ & $ 1.67$ & $-0.02$ & $ 1.17$ & $0.41$ \\
\hline
$3.05\,N$  &  $0.30 $ & $-0.45$ & $ 0.58$ & $-0.40$ & $ 0.03$ & $0.16$ \\
$6.10\,N$  &  $0.67 $ & $-0.20$ & $ 0.30$ & $-0.37$ & $ 0.15$ & $0.08$ \\
$9.15\,N$  &  $0.50 $ & $-0.46$ & $ 0.16$ & $-0.57$ & $-0.14$ & $0.22$ \\
$12.20\,N$ &  $0.79 $ & $-0.40$ & $-0.12$ & $-0.41$ & $-0.22$ & $0.01$ \\
$15.25\,N$ &  $0.08 $ & $-0.73$ & $-0.52$ & $-0.79$ & $-0.32$ & $0.20$ \\
$18.30\,N$ &  $0.18 $ & $-0.01$ & $ 0.07$ & $ 0.03$ & $ 0.08$ & $0.62$ \\
$24.40\,N$ &  $0.57 $ & $ 0.55$ & $ 0.21$ & $-0.02$ & $ 1.05$ & $0.19$ \\

\hline
\end{tabular}
\begin{list} {Table Notes.}
\item  Residuals correspond to the observed $-$ synthesized values.
\end{list}
\label{residualw}
\end{table}

%
%********************************************************************************
%
%**********************************************************************************
%
\renewcommand{\tabcolsep}{1.9mm}
\begin{table}
\caption{Continuum residuals}
\footnotesize
\begin{tabular} {l|r r r r r r}
%\multicolumn{7}{c}{Continuum residuals} \\
\hline
\hline
R (\arcsec) & $\lambda5300$ & $\lambda5313$ &
$\lambda5546$ & $\lambda5800$ & $\lambda5822$ & $\lambda6630$ \\
\hline
$0.00$     & $0.01$ &  $0.08$ & $ 0.01$ & $ 0.02$ & $ 0.02$ & $ 0.04$ \\
$3.05\,S$  & $0.04$ &  $0.05$ & $-0.02$ & $ 0.01$ & $ 0.00$ & $ 0.01$ \\
$6.10\,S$  & $0.08$ &  $0.08$ & $ 0.00$ & $ 0.02$ & $ 0.02$ & $ 0.01$ \\
$9.15\,S$  & $0.06$ &  $0.07$ & $-0.02$ & $ 0.00$ & $ 0.00$ & $-0.01$ \\
$12.20\,S$ & $0.06$ &  $0.06$ & $-0.01$ & $-0.01$ & $-0.02$ & $ 0.00$ \\
$15.25\,S$ & $0.08$ &  $0.08$ & $ 0.02$ & $-0.01$ & $ 0.00$ & $-0.02$ \\
$18.30\,S$ & $0.06$ &  $0.07$ & $ 0.00$ & $ 0.03$ & $ 0.01$ & $ 0.02$ \\
$24.40\,S$ & $0.06$ &  $0.07$ & $ 0.02$ & $ 0.02$ & $ 0.01$ & $ 0.00$ \\
\hline
$3.05\,N$  & $0.04$ &  $0.04$ & $-0.03$ & $ 0.00$ & $ 0.01$ & $ 0.01$ \\
$6.10\,N$  & $0.04$ &  $0.05$ & $-0.03$ & $ 0.01$ & $ 0.01$ & $ 0.00$ \\
$9.15\,N$  & $0.05$ &  $0.05$ & $-0.01$ & $ 0.01$ & $-0.01$ & $ 0.00$ \\
$12.20\,N$ & $0.05$ &  $0.05$ & $-0.01$ & $ 0.00$ & $ 0.01$ & $ 0.01$ \\
$15.25\,N$ & $0.06$ &  $0.06$ & $ 0.00$ & $ 0.02$ & $ 0.00$ & $ 0.01$ \\
$18.30\,N$ & $0.04$ &  $0.04$ & $-0.01$ & $ 0.01$ & $-0.01$ & $ 0.00$ \\
$24.40\,N$ & $0.04$ &  $0.04$ & $-0.01$ & $ 0.01$ & $ 0.00$ & $ 0.00$ \\

\hline
\end{tabular}
\begin{list} {Table Notes.}
\item Residuals correspond to the observed $-$ synthesized values.
\end{list}

\label{residualc}
\end{table}
%
%**********************************************************************************
%
\subsection{Integrated Colour Index }

\begin{figure}
\resizebox{\hsize}{!}{\includegraphics{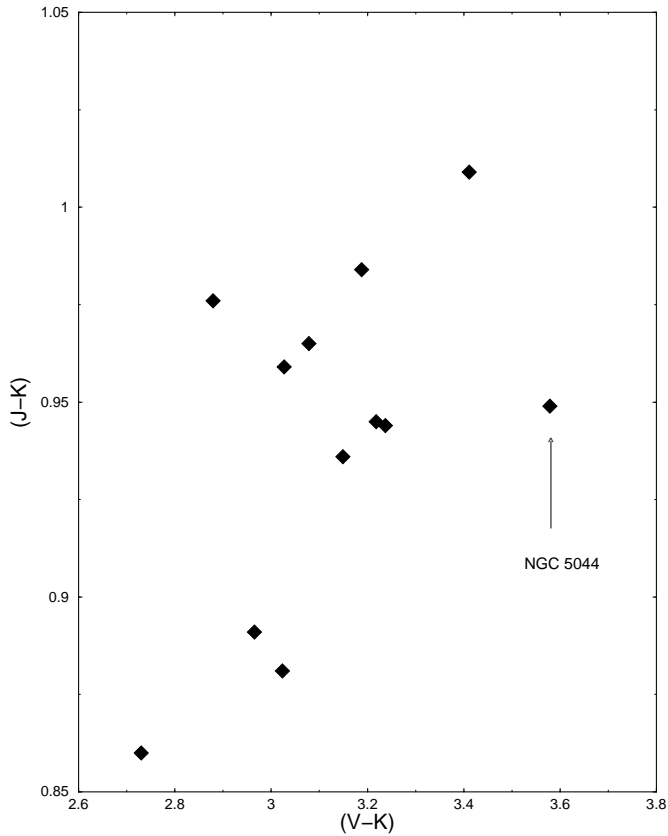}}
\caption{Colour-colour diagram (J-K) vs. (V-K) }
\label{fig8}
\end{figure}

In order to put NGC\,5044 in the context of other elliptical galaxies rich in interestellar 
medium, we build the colour-colour diagram (J-K) vs. (V-K), using the J, K and V photometric 
data for a sample of elliptical galaxies from Rembold et al. (2002) and  NED. The results 
are presented in Fig.~\ref{fig8}, showing that NGC\,5044 has the reddest (V-K) colour. 
This result is probably due to the fact that besides the old stellar population contribution 
to the (V-K) colour index, NGC\,5044 also presents mid-IR (6.7 to 15$\mu\,m$) dust emission 
(Ferrari et al. 2002).

\section{Ionized gas}

Emission gas has been detected in a large number of elliptical galaxies. However, the origin
of this gas and the ionization source are not yet conclusively established. NGC\,5044 presents
conspicuous emission lines throughout the its body. The mass of ionized gas in NGC\,5044 is 
estimated between $10^3$ and $10^5\,M_\odot$ (Macchetto et al. 1996). In this section we 
investigate the physical conditions and ionization source of the gas in NGC\,5044 using the 
results of the stellar population synthesis.

The properties of the emission gas are derived from line fluxes, measured  in the stellar
population subtracted spectra (Section~5). The emission lines used in the analysis are:
\ion{H}{$\alpha$}, \ion{[N}{ii]}$_{\lambda6548}$, \ion{[N}{ii]}$_{\lambda6584}$,
\ion{[S}{ii]}$_{\lambda6717}$ and \ion{[S}{ii]}$_{\lambda6731}$. In Tab.~\ref{tableNIISII} we 
list, for each extraction region, the absolute flux, the flux relative to that of \ion{H}{$\alpha$} 
in the central region, and the FWHM for the above emission lines. The uncertainty attributed to 
each measured flux is based on a gaussian least-squares fit.

\renewcommand{\tabcolsep}{1.55mm}
\begin{table}
\caption{Emission line parameters.}
\tiny
\begin{tabular} {l|c c c c}
\hline
\hline
$R (\arcsec) $ & $lines$ & $F_{\lambda}$ &
$\frac{F_\lambda}{F_{H\alpha\,(R=0)}}$ & $FWHM$ \\

& & $(10^{-16}\,erg\,s^{-1}\,cm^{-2})$ & & (\AA) \\

\hline

           & $H\alpha$  & $6.80 \pm 0.23$ & $1.00$ & $13.62$ \\
           & $[NII]_{\lambda6548}$ & $22.0 \pm 0.75$ & $3.23 \pm 0.15$ & $21.27$ \\
 $0.00$  & $[NII]_{\lambda6584}$ & $7.33 \pm 0.25$ & $1.07 \pm 0.05$ & $21.27$ \\
           & $[SII]_{\lambda6717}$ & $8.73 \pm 0.16$ & $1.28 \pm 0.04$ & $23.64$ \\
           & $[SII]_{\lambda6731}$ & $5.83 \pm 0.11$ & $0.85 \pm 0.03$ & $23.64$ \\
 \hline
           & $H\alpha  $           & $4.29 \pm 0.11$ & $0.63 \pm 0.03$ & $11.82$ \\
           & $[NII]_{\lambda6548}$ & $7.34 \pm 0.19$ & $1.08 \pm 0.04$ & $11.21$ \\
 $3.05S$  & $[NII]_{\lambda6584}$ & $2.44 \pm 0.06$ & $0.35 \pm 0.01$ & $11.21$ \\
           & $[SII]_{\lambda6717}$ & $2.58 \pm 0.08$ & $0.37 \pm 0.02$ & $11.10$ \\
           & $[SII]_{\lambda6731}$ & $1.96 \pm 0.06$ & $0.28 \pm 0.01$ & $11.10$ \\
\hline
           & $H\alpha  $           & $3.27 \pm 0.14$ & $0.48 \pm 0.03$ & $11.78$ \\
           & $[NII]_{\lambda6548}$ & $5.91 \pm 0.25$ & $0.86 \pm 0.04$ & $12.24$ \\
$6.10S$   & $[NII]_{\lambda6584}$ & $1.97 \pm 0.08$ & $0.28 \pm 0.01$ & $12.24$ \\
           & $[SII]_{\lambda6717}$ & $2.04 \pm 0.10$ & $0.30 \pm 0.02$ & $10.73$ \\
           & $[SII]_{\lambda6731}$ & $1.42 \pm 0.07$ & $0.21 \pm 0.01$ & $10.73$ \\
\hline
           & $H\alpha $            & $2.91 \pm 0.11$ & $0.43 \pm 0.02$ & $15.08$ \\
           & $[NII]_{\lambda6548}$ & $4.36 \pm 0.17$ & $0.64 \pm 0.03$ & $12.19$ \\
$9.15S$   & $[NII]_{\lambda6584}$ & $1.45 \pm 0.05$ & $0.21 \pm 0.01$ & $12.19$ \\
           & $[SII]_{\lambda6717}$ & $1.50 \pm 0.08$ & $0.22 \pm 0.01$ & $12.12$ \\
           & $[SII]_{\lambda6731}$ & $1.24 \pm 0.07$ & $0.18 \pm 0.01$ & $12.12$ \\
\hline
           & $H\alpha  $           & $2.49 \pm 0.09$ & $0.37 \pm 0.02$ & $18.32$ \\
           & $[NII]_{\lambda6548}$ & $2.36 \pm 0.08$ & $0.34 \pm 0.01$ & $12.59$ \\
$12.20S$  & $[NII]_{\lambda6584}$ & $0.78 \pm 0.03$ & $0.11 \pm 0.00$ & $12.59$ \\
           & $[SII]_{\lambda6717}$ & $0.60 \pm 0.05$ & $0.08 \pm 0.01$ & $11.12$ \\
           & $[SII]_{\lambda6731}$ & $0.57 \pm 0.04$ & $0.08 \pm 0.00$ & $11.12$ \\
\hline
           & $H\alpha  $           & $3.39 \pm 0.20$ & $0.49 \pm 0.03$ & $32.46$ \\
           & $[NII]_{\lambda6548}$ & $2.41 \pm 0.14$ & $0.35 \pm 0.02$ & $10.30$ \\
$15.25S$  & $[NII]_{\lambda6584}$ & $0.80 \pm 0.05$ & $0.11 \pm 0.01$ & $10.30$ \\
           & $[SII]_{\lambda6717}$ & $0.63 \pm 0.03$ & $0.09 \pm 0.00$ & $12.84$ \\
           & $[SII]_{\lambda6731}$ & $1.15 \pm 0.05$ & $0.17 \pm 0.01$ & $12.84$ \\
\hline
           & $H\alpha  $           & $3.43 \pm 0.13$ & $0.50 \pm 0.02$ & $12.55$ \\
           & $[NII]_{\lambda6548}$ & $5.47 \pm 0.21$ & $0.80 \pm 0.04$ & $11.88$ \\
$18.30S$  & $[NII]_{\lambda6584}$ & $1.82 \pm 0.07$ & $0.26 \pm 0.01$ & $11.88$ \\
           & $[SII]_{\lambda6717}$ & $1.00 \pm 0.15$ & $0.14 \pm 0.02$ & $10.26$ \\
           & $[SII]_{\lambda6731}$ & $1.15 \pm 0.17$ & $0.17 \pm 0.02$ & $10.26$ \\
\hline
           & $H\alpha  $           & $4.06 \pm 0.23$ & $0.60 \pm 0.04$ & $17.73$ \\
           & $[NII]_{\lambda6548}$ & $4.88 \pm 0.27$ & $0.71 \pm 0.05$ & $13.16$ \\
$24.40S$  & $[NII]_{\lambda6584}$ & $1.62 \pm 0.09$ & $0.24 \pm 0.01$ & $13.16$ \\
           & $[SII]_{\lambda6717}$ & $1.58 \pm 0.21$ & $0.23 \pm 0.03$ & $12.98$ \\
           & $[SII]_{\lambda6731}$ & $1.09 \pm 0.14$ & $0.16 \pm 0.02$ & $12.98$ \\
\hline
\hline

           & $H\alpha $            & $4.73 \pm 0.38$ & $1.13 \pm 0.06$ & $11.08$ \\
           & $[NII]_{\lambda6548}$ & $7.84 \pm 0.62$ & $1.15 \pm 0.10$ & $11.85$ \\
  $3.05N$ & $[NII]_{\lambda6584}$ & $2.61 \pm 0.21$ & $0.38 \pm 0.03$ & $11.85$ \\
           & $[SII]_{\lambda6717}$ & $2.32 \pm 0.45$ & $0.34 \pm 0.06$ & $10.73$ \\
           & $[SII]_{\lambda6731}$ & $1.63 \pm 0.32$ & $0.24 \pm 0.05$ & $10.73$ \\
 \hline
           & $H\alpha $            & $3.74 \pm 0.14$ & $0.55 \pm 0.03$ & $11.08$ \\
           & $[NII]_{\lambda6548}$ & $4.56 \pm 0.23$ & $0.67 \pm 0.04$ & $12.11$ \\
 $6.10N$  & $[NII]_{\lambda6584}$ & $1.52 \pm 0.08$ & $0.22 \pm 0.03$ & $12.11$ \\
           & $[SII]_{\lambda6717}$ & $0.96 \pm 0.07$ & $0.14 \pm 0.01$ & $10.61$ \\
           & $[SII]_{\lambda6731}$ & $0.70 \pm 0.05$ & $0.10 \pm 0.01$ & $10.61$ \\
 \hline
           & $H\alpha $            & $2.08 \pm 0.11$ & $0.30 \pm 0.02$ & $11.32$ \\
           & $[NII]_{\lambda6548}$ & $3.51 \pm 0.19$ & $0.51 \pm 0.03$ & $12.80$ \\
 $9.15N$  & $[NII]_{\lambda6584}$ & $1.17 \pm 0.06$ & $0.17 \pm 0.01$ & $12.80$ \\
           & $[SII]_{\lambda6717}$ & $0.72 \pm 0.06$ & $0.10 \pm 0.01$ & $11.35$ \\
           & $[SII]_{\lambda6731}$ & $0.63 \pm 0.05$ & $0.09 \pm 0.00$ & $11.35$ \\
 \hline
           & $H\alpha $            & $1.25 \pm 0.10$ & $0.18 \pm 0.01$ & $11.74$ \\
           & $[NII]_{\lambda6548}$ & $2.65 \pm 0.21$ & $0.39 \pm 0.03$ & $15.85$ \\
 $12.20N$ & $[NII]_{\lambda6584}$ & $0.88 \pm 0.77$ & $0.13 \pm 0.01$ & $15.85$ \\
           & $[SII]_{\lambda6717}$ & $0.50 \pm 0.02$ & $0.07 \pm 0.00$ & $11.00$ \\
           & $[SII]_{\lambda6731}$ & $0.40 \pm 0.01$ & $0.06 \pm 0.00$ & $11.00$ \\
 \hline
           & $H\alpha $            & $3.51 \pm 0.17$ & $0.51 \pm 0.03$ & $33.26$ \\
           & $[NII]_{\lambda6548}$ & $2.41 \pm 0.12$ & $0.35 \pm 0.02$ & $10.31$ \\
 $15.25N$ & $[NII]_{\lambda6584}$ & $0.80 \pm 0.04$ & $0.12 \pm 0.00$ & $10.31$ \\
           & $[SII]_{\lambda6717}$ & $0.40 \pm 0.09$ & $0.06 \pm 0.01$ & $10.89$ \\
           & $[SII]_{\lambda6731}$ & $0.37 \pm 0.08$ & $0.05 \pm 0.01$ & $10.89$ \\
 \hline
           & $H\alpha $            & $1.32 \pm 0.09$ & $0.19 \pm 0.01$ & $11.45$ \\
           & $[NII]_{\lambda6548}$ & $1.93 \pm 0.13$ & $0.28 \pm 0.02$ & $11.88$ \\
 $18.30N$ & $[NII]_{\lambda6584}$ & $0.64 \pm 9.22$ & $0.09 \pm 1.35$ & $11.88$ \\
           & $[SII]_{\lambda6717}$ & $0.54 \pm 0.14$ & $0.08 \pm 0.02$ & $8.76 $ \\
           & $[SII]_{\lambda6731}$ & $0.48 \pm 0.13$ & $0.07 \pm 0.02$ & $8.76 $ \\
 \hline
           & $H\alpha $            & $3.98 \pm 0.22$ & $0.58 \pm 0.03$ & $11.66$ \\
           & $[NII]_{\lambda6548}$ & $4.82 \pm 0.27$ & $0.71 \pm 0.05$ & $10.71$ \\
 $24.40N$ & $[NII]_{\lambda6584}$ & $1.60 \pm 0.90$ & $0.23 \pm 0.13$ & $10.71$ \\
           & $[SII]_{\lambda6717}$ & $1.60 \pm 0.13$ & $0.23 \pm 0.02$ & $11.87$ \\
           & $[SII]_{\lambda6731}$ & $1.95 \pm 0.15$ & $0.28 \pm 0.02$ & $11.87$ \\

\hline
\end{tabular}
\begin{list} {Table Notes.}
\item Column~4: Flux normalized to the central region $F_{H\alpha}$ ; Column~5: FWHM according 
to a gaussian fit.

\end{list}
\label{tableNIISII}
\end{table}

Assuming that the emission lines are formed only by recombination, the number of
ionizing photons (Q(H)) can be  calculated as
\begin{eqnarray}
Q(H) =\frac{L_{H\alpha}}{h\nu_\alpha}\frac{\alpha_B(H^0,T)}{\alpha_{H\alpha} (H^0,T)},
\label{Q}
\end{eqnarray}
where the \ion{H}{$\alpha$} luminosity is $L_{H\alpha} = 4\pi D^2F_{H\alpha}$, $\alpha_B(H^0,T)$ 
is the recombination coefficient summed over all energy levels, and $\alpha_{H\alpha}(H^0,T)$
is the recombination coefficient at \ion{H}{$\alpha$}.  With a Hubble constant of
$H_0 = 75\rm\,km\,s^{-1}\,Mpc^{-1}$, the distance to NGC\,5044 is $D=36\,\rm Mpc$.
$L_{H\alpha}$ and Q(H) values are given in Tab~\ref{6.3}.

\begin{figure}
\resizebox{\hsize}{!}{\includegraphics{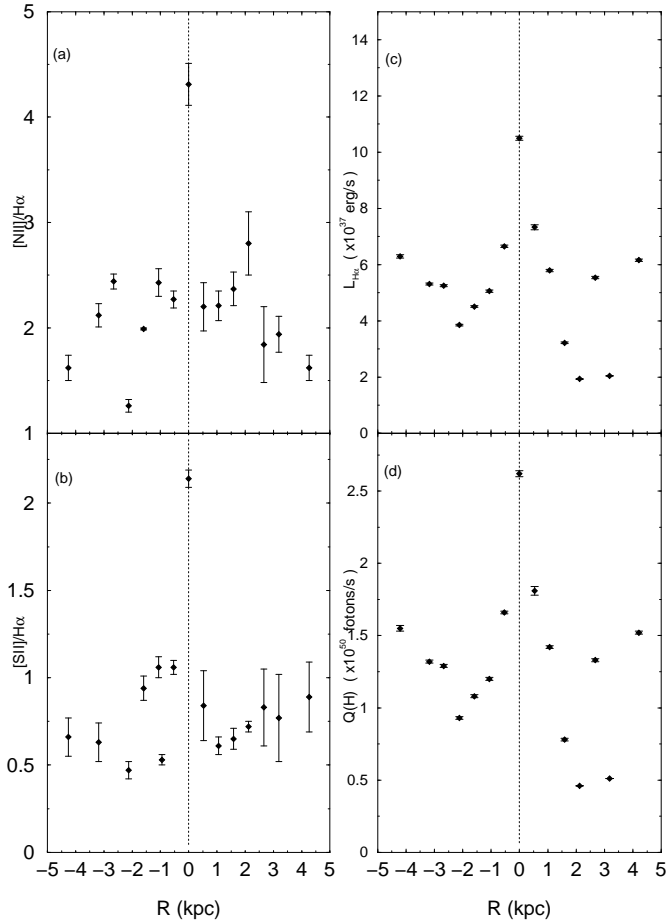}}
\caption{Spatial distribution of line-ratios and Q(H).}
\label{fig9}
\end{figure}

In order to infer on the spatial properties of the emission gas, we plot in Fig.~\ref{fig9}
the line ratios  $\frac{\ion{[N}{ii]}_{\lambda\lambda 6548, 84}}{H\alpha}$ (panel (a)),
$\frac{\ion{[S}{ii]}_{\lambda\lambda 6717, 31}}{H\alpha}$ (panel (b)), $L_{H\alpha}$ (panel (c)), and
Q(H) (panel (d)).

%********************************************************************************************************
\renewcommand{\tabcolsep}{4.5mm}
\begin{table}
\caption{Luminosity and ionizing photons}
\begin{tabular} {l c c}
\hline
\hline
 R(\arcsec) & $L_{H\alpha}$ & $Q(H)$ \\
 & ($\times 10^{37} \,erg\,s^{-1}$) & ($\times 10^{50}\,photons\,s^{-1}$) \\
 \hline
$0.00$  & $10.5 \pm 0.07$ & $2.62 \pm 0.02$  \\
$3.05S$ & $6.65 \pm 0.04$ & $1.66 \pm 0.01$  \\

$6.10S$ & $5.06 \pm 0.04$ & $1.20 \pm 0.01$  \\
$9.15S$ & $4.51 \pm 0.03$ & $1.08 \pm 0.01$  \\
$12.20S$ & $3.85 \pm 0.03$ & $0.93 \pm 0.01$  \\
$15.25S$ & $5.25 \pm 0.04$ & $1.29 \pm 0.01$  \\
$18.30S$ & $5.31 \pm 0.03$ & $1.32 \pm 0.01$  \\
$24.40S$ & $6.29 \pm 0.06$ & $1.55 \pm 0.02$  \\
\hline
$3.05N$ & $7.33 \pm 0.09$ & $1.81 \pm 0.03$  \\
$6.10N$ & $5.79 \pm 0.04$ & $1.42 \pm 0.01$  \\
$9.15N$ & $3.22 \pm 0.03$ & $0.78 \pm 0.01$  \\
$12.20N$ & $1.93 \pm 0.01$ & $0.46 \pm 0.00$  \\
$15.25N$ & $5.54 \pm 0.04$ & $1.33 \pm 0.01$  \\
$18.30N$ & $2.04 \pm 0.01$ & $0.51 \pm 0.00$  \\
$24.40N$ & $6.16 \pm 0.04$ & $1.52 \pm 0.01$  \\
\hline
\end{tabular}
%\begin{list} {Table Notes.}
%\item Column 1: S and N indicate south and north extractions, respectively; Column 5:
%internal reddening. to the Galactic law.
%\end{list}
\label{6.3}
\end{table}
%
%**********************************************************************************
%**********************************************************************************
%

In the central region, NGC\,5044 presents $\frac{\ion{[N}{ii]}}{H\alpha}=4.2$ and
$\frac{\ion{[S}{ii]}}{H\alpha}>1$.  Notice that the ratio
$\frac{\ion{[N}{ii]}}{H\alpha}$ is larger than 1.0  for all sampled regions (panel (a)).

In order to investigate the nature of the ionized source in the external regions of NGC\,5044,
we test two scenarios:

%       {\it (i)}    the gas is ionized by single O5 star.  The number of ionizing %photons produced by
%one  O5 star with effective temperature of 30000\,K, using Kurucz's (1979) model, is
%$1.76 \times 10^{48}\,\rm photons\,s^{-1}$;  however the number of ionizing photons %measured in
%the region at $3.05\arcsec$ is $\sim 1.62 \times 10^{50}\,\rm photons\,s^{-1}$ (see %Tab.~\ref{6.3})

{\it (i)}  the gas is ionized by an \ion{H}{ii} region.  In order to reproduce the number of
ionizing photons observed in the region at $3.05\arcsec$, we have built an \ion{H}{ii} region 
according to a Salpeter initial mass function. The \ion{H}{ii} region model which best reproduces 
the observed number of ionizing photons is listed in Tab.~\ref{estrelas}, where we give the 
distribution of stars according to spectral type and mass; in the last column we give the number 
of $\ion{Ly}{\alpha}$ photons produced by each star (Dottori 1991). This model \ion{H}{ii} region 
produces $\sim2.59\times 10^{50}\,\rm photons\,s^{-1}$ which is comparable to the observed value.  
However, such an \ion{H}{ii} region in NGC\,5044 would be bright enough to be photometrically 
detected in visible images obtained with the ESO 3.60302fig10.eps0302fig10.eps\,m telescope. The V-band image does not 
present any signature of such an \ion{H}{ii} region.

\begin{figure}
\resizebox{\hsize}{!}{\includegraphics{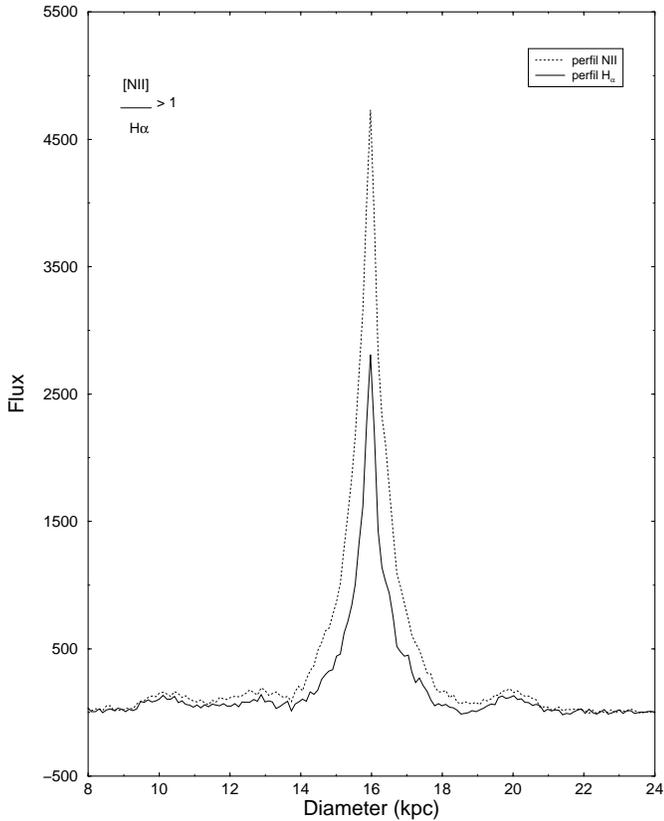}}
\caption{Spatial profiles of \ion{H}{$\alpha$} and [NII]. Notice that ${\rm [NII]/H_\alpha > 1}$ 
along the galaxy.}
\label{fig10}
\end{figure}

{\it (ii)} the gas is ionized by a hot, post-AGB star. The number of ionizing photons produced by 
one post-AGB star with effective temperature of $\sim 150\,000\,K$ corresponds to
$\sim6.11\times 10^{50}\,\rm photons\,s^{-1}$, about 5 times the number of photons observed in 
the region at $3.05\arcsec$. The spatial profiles of $[\ion{N}{ii}]$ and \ion{H}{$\alpha$} 
extracted along the slit, shown in Fig.~\ref{fig10}, reveal that $\frac{[NII]}{H\alpha}>2$. 
However, according to the ionization models for old stars in elliptical galaxies (Binette et al. 
1994) the ratio $\frac{\ion{N[}{ii]}}{\ion{H}{\alpha}}$ measured in NGC\,5044 could only be 
produced by a much larger Q(H) than the one we estimate, in particular the high value measured 
in the central region. Binette et al.'s (1994) models do not reproduce as well the observed  
ratio $\frac{[SII]_{6731}}{H\alpha}$. A high ratio $\frac{\ion{N[}{ii]}}{\ion{H}{\alpha}}$ is 
characteristic of AGNs.

In conclusion we find that in the central region the ionization source must be a AGN, while in 
the external region the lower ratio $\frac{\ion{N[}{ii]}}{\ion{H}{\alpha}}\sim2$ can be 
accounted for by post-AGB stars.

%
%**********************************************************************************
%**********************************************************************************
%

\renewcommand{\tabcolsep}{3.0mm}
\begin{table}
\caption{$\ion{H}{ii}$ region model} 
\begin{tabular} {l c c c}
\hline
\hline
{\it Spectral type} & $\frac{M}{M_\odot}$ & {\it Number} &
$N_{\ion{Ly}{\alpha}}$ \\
 & & & ($\times 10^{48}\,photons\,s^{-1}$) \\
\hline
$O5$ & $30$ & $20 $ & $2.00  $ \\
     & $22$ & $41 $ & $0.25  $ \\       
$B0$ & $17$ & $75 $ & $0.082 $ \\
     & $15$ & $101$ & $0.043 $ \\
     & $13$ & $142$ & $0.015 $ \\
     & $10$ & $264$ & $0.0021$ \\
\hline
\end{tabular}
\begin{list} {Table Notes.}
\item Column 3: Number of stars of each spectral type.
\end{list}   
\label{estrelas}
\end{table}

%****************************** CONCLUS�S ****************************************

\section{Conclusions}

        The stellar population, metallicity distribution and ionized gas in NGC\,5044 have been 
investigated in this paper by means of long-slit spectroscopy and stellar population synthesis.  
With respect to the spatial distribution of metal-strength indices (corrected for velocity 
dispersion), the \ion{Mg}{2} profile increases towards the central region, contrary to what is 
observed for the profile of \ion{Fe}{i}$_{5270}$. This difference in slopes may be accounted for 
by an enhancement of $\alpha$-elements in general, Mg in particular. The \ion{Mg}{2} 
gradient slope compared with the mass of NGC\,5044 indicates 
a monolitic collapse.  However, the measure [Fe/H] gradient slope is not characteristic of 
one monolitic collapse. 

The stellar population synthesis gives that most metallic component ($[Z/Z_\odot]\sim0.0$) dominates the 
$\lambda5870$\,\AA\ flux in the central region of NGC\,5044. In the nucleus, this component 
contributes with $\sim42\%$ of the total flux, while in the external regions the contribution decreases 
to $\sim8.0\%$.
The component with $[Z/Z_\odot]\sim-0.4$ contributes with $\sim32\%$ in the central 
region, and $\sim55\,\%$ in the external regions. The metal-poor component
($[Z/Z_\odot]\sim-1.1$) contributes with $\sim26\,\%$ in the central region, and $\sim 37\,\%$
in the external regions. The spatial distribution of the internal reddening E(B$-$V)$_{\rm{i}}$ 
along the north and south directions does not characterize a gradient.

        The large values of the ratio $\frac{\ion{[N}{ii]}}{H\alpha}$ observed in all sampled regions 
characterizes the presence of a non-thermal ionization source, such as a low-luminosity AGN and/or shock 
ionization.  However, in the external regions an additional ionization source is necessary to explain 
the emission lines, which might be hot, post-AGB stars.

%*********************************** REFERENCE ************************************

%**********************************************************************************
\end{document}